\documentclass[11pt]{article}

\usepackage[english]{babel}

\usepackage[utf8]{inputenc} 

\usepackage[utf8]{inputenc}
\usepackage{textcomp}

\usepackage{geometry} 
\usepackage{libertine} 

\usepackage{color}
\usepackage{amsmath}
\usepackage[autostyle]{csquotes}
\usepackage{textcomp}
\usepackage{amssymb} %
\usepackage{url}
\usepackage{hyperref}
\usepackage{graphicx} 
\graphicspath{{figures/}} 
\usepackage{cite}
\usepackage{float}
\usepackage{todonotes} 
\usepackage{comment} %
\usepackage{epigraph} %
\usepackage{cleveref} %
\usepackage{listings}

\newcommand{\myref}[1]{\cref{#1}} %
\newcommand{\Myref}[1]{\Cref{#1}} %
\newcommand{\sparql}{SPARQL}

\newcommand{\schemadule}[1]{\textbf{S}:\quad #1} %
\newcommand{\emdash}{\,\textemdash{}\,}

\hypersetup{
 pdfauthor={},
 pdftitle={},
 pdfkeywords={},
 pdfsubject={},
 pdflang={English}} %

\begin{document} %

\begin{flushleft}
\begin{tabular}{p{11.1cm}r}
Monografias em Ci\^{e}ncia da Computa\c{c}\~{a}o, No. XX/AA & ISSN: 0103-9741 \\
Editor: Prof. Carlos Jos\'{e} Pereira de Lucena   & Mes por extenso, AAAA
\end{tabular}
\end{flushleft}

\LARGE

\bigskip

\begin{center}
{\bf NoSQL Graph Databases: an overview}
\end{center}

\normalsize
\bigskip

\begin{center}
{\bf Veronica Santos, Bruno Cuconato}
\end{center}
\begin{center}
Departamento de Informática, PUC-Rio
\end{center}

\begin{center}
   \{vdsantos, bclaro\}@inf.puc-rio.br
\end{center}

\date{}

$~$ \\

\thispagestyle{empty}

\noindent {\bf Abstract.}
Graphs are the most suitable structures for modeling objects and interactions in applications where component inter-connectivity is a key feature. %
There has been increased interest in graphs to represent domains such as social networks, web site link structures, and biology. %
Graph stores recently rose to prominence along the NoSQL movement. %
In this work we will focus on NOSQL graph databases, describing their peculiarities that sets them apart from other data storage and management solutions, and how they differ among themselves. %
We will also analyze in-depth two different graph database management systems — AllegroGraph and Neo4j that uses the most popular graph models used by NoSQL stores in practice: the resource description framework (RDF) and the labeled property graph (LPG), respectively. %

\medskip

\medskip


\bigskip

\noindent {\bf Resumo.}

\medskip

\medskip

\noindent {\bf Palavras-chave:} chave1, chave2, chave3 ... \\

\newpage
\pagenumbering{roman} \setcounter{page}{2}

\,\,\vspace*{17cm}

\begin{flushleft}
\textbf{In charge for publications:} \\
%
Rosane Teles Lins Castilho \\
Assessoria de Biblioteca, Documenta\c{c}\~{a}o e Informa\c{c}\~{a}o \\
PUC-Rio Departamento de Inform\'{a}tica \\ Rua Marqu\^{e}s de
S\~{a}o Vicente, 225 - G\'{a}vea  \\
22451-900 Rio de Janeiro RJ Brasil \\
Tel. +55 21 3527-1500 \\
E-mail: bib-di@inf.puc-rio.br  \\
Web site: http://bib-di.inf.puc-rio.br/techreports/ \\
\end{flushleft}


\newpage
\pagenumbering{arabic} \setcounter{page}{1}

\section{Introduction}
\label{sec:intro} %

\epigraph{In an extreme view, the world can be seen as only connections, nothing else. We think of a dictionary as the repository of meaning, but it defines words only in terms of other words. I liked the idea that a piece of information is really defined only by what it's related to, and how it's related. There really is little else to meaning. The structure is everything. There are billions of neurons in our brains, but what are neurons? Just cells. The brain has no knowledge until connections are made between neurons. All that we know, all that we are, comes from the way our neurons are connected.}{Tim Berners-Lee}


Graph stores recently rose to prominence along the NoSQL movement. %
As such, their use was stimulated by the ubiquitous presence of the web and of mobile devices, which caused an explosion in the availability of data of all sorts. %
The new applications of web technologies demanded new database requirements, whose support in traditional database management systems at the time was not ideal \cite{davoudian2018survey}. %
New database management systems could not only make different choices in the trade-offs imposed by the CAP theorem, but could do so without worrying about legacy code and previous applications. %

Although graph stores piggybacked on the NoSQL movement, they also have motives to prominence of their own. %
Besides being well-understood and having several applications, graphs also enjoy the availability of efficient algorithms for search and traversal, which makes for relatively easy and fast analysis of data encoded as graphs. %
Even though graph theory is old and well-known, new applications of it are discovered everyday. %
Major companies (in both net worth and in brand recognition) have graphs as the theoretical background of their core services, such as Google (PageRank algorithm for ranking search results) and Facebook (the social graph). %
There has been increased interest in graphs to represent domains such as social networks, web site link structures, and biology. %
There are many uses for graphs since graphs are the most suitable structures for modeling objects and interactions in applications where component inter-connectivity is a key feature\cite{angles2008survey}. %
For example: in biology, graphs can be used to represent metabolic networks, protein-protein interaction networks, chemical structure graphs, gene clusters, and genetic maps. %

Another domain where graphs excel is that of data provenance. %
Data provenance is the lineage of data item, which describes what it is and how it came to be, including data about creation/transformation process and its input data. %
A directed acyclic graph (DAG) is a suitable data structure to store provenance of data item based on data relationships. %
A common question in provenance systems is: if a data object (starting node) is determined to be incorrect what are the other data objects (nodes) derived from or affected by it? %
This type of question is answered with a traversal query in the DAG\cite{comparison_provenance} using algorithms like breadth-search first or depth-search first. %

The rest of the paper is organized as follows. %
In section 2, we will give the theoretical foundations for graph models and graph databases. %
Section 3 is dedicated to overview a few of the studies on graph databases found in the literature and to overview NoSQL Graph Database Systems frequently mentioned in those works. %
Section 4 we will analyze in-depth two different graph database management systems — AllegroGraph and Neo4j that uses the most popular graph models in practice: the resource description framework (RDF) and the labeled property graph (LPG), respectively. %
The last section we used to make some final considerations about this two systems. %

\section{Theoretical Foundations}
\label{sec:definition}
The rise of the Web 2.0 made possible new kinds of applications and instigated new development processes that were either incompatible or limited by the relational DBMSs of the time. %
Since then most DBMSs have adapted and expanded their use cases, but in the meantime a wealth of data storage and management systems have been created. %
These new ``data stores'' are collectively dubbed ‘NoSQL’. %
Relational database technologies support schema-full data models with integrity constraints and NoSQL databases offer more flexible data models, that can be schema-less, in order to reduce the cost of schema evolution. %
Using a schema-less data model data will be interpreted at the application level (schema-on-read) since stored data can have arbitrary structures as they are not explicitly, uniquely and \textit{a priori} defined by data definition language (schema-on-write). %
NoSQL stores share design principles like allowing flexible data schemas (or even no schemas at all) and relaxing some transaction guarantees for improved performance, but also differ significantly between themselves, most noticeably in their different data models. %
Because of this diversity is is difficult give a clear-cut definition for NoSQL stores; they are usually defined by a mixture of opposing them to traditional relational systems and of defining a common set of properties that they are supposed to share, as done by \cite{davoudian2018survey} and echoed here. %

Data model, as stated in \cite{davoudian2018survey}, specifies how real-world entities and their relationships are represented and operated. %
As noted earlier, NoSQL database systems are often distinguished by the data models they use, which are the following: key-value, wide-column, document, or graph models. %
A fifth category named multi-model can be used to classify some database management systems like ArangoDB, in that they support more than one of these data models. %
In this work we will focus on NOSQL graph databases, describing their peculiarities that sets them apart from other data storage and management solutions, and how they differ among themselves. %
\todo [disable]{this might be useful in the intro itself; let's delete this comment only when we are about to close this work and can check if we fulfill these goals\\ %
>br: yeah, I don't think we did that exactly… >vds talvez sim, sem tanta profundidade, coloquei no abstract} %

We start by defining graph stores\footnote{We use `graph stores' and `graph databases' interchangeably, like \cite{davoudian2018survey} but unlike \cite{angles2008survey}.} themselves. %
Any database system that 
exposes a graph data model through CRUD operations is be classified as a graph database. %
When inserting, manipulating, and deleting data in their stores, users think in terms of vertices and edges\emdash{}they have a mental graph model of their data; query and manipulation languages speak of the same entities; and graph algorithms inform their implementations. %
\cite{angles2008survey} defines graph database models as being “[…] those in which data structures for the schema and instances are modeled as graphs or generalizations of them, and data manipulation is expressed by graph-oriented operations and type constructors.” %

\subsection{Graph Models}
\label{sec:graph-models} %

The mathematical definition of a simple graph \(G\) is a tuple \((V, E)\), where \(V\) is its set of vertices, and \(E \subseteq{} E \times{} E\) is its set of edges. %
In case we want to have an undirected graph, an edge \(e = {v, w} \in E\) is simply a set of two vertices, whereas if we want to have a directed graph we take \(e\) to be a tuple whose first element is the source and the second element is the target vertex. %

The definition of a simple graph can be extended to model more complex kinds of graphs. %
A weighted graph can be modeled by a triple \(V, E, w\), where \(w : E \mapsto{} \mathbb{R}\) is function that gives the weight of a given edge.%
\footnote{Another option is to include the weight in the edges themselves, so that in undirected graphs we would have an edge \(e = ({v,w}, w)\) were \({v, w}\) is a set of two vertices and \(w \in \mathbb{R}\) is the weight of the edge. %
  In directed graphs the edges would become triples in the set \(V \times{} V \times{} \mathbb{R}\).} %
A hypergraph \(H\) generalizes a simple graph \(G\) allowing any of its edges to be a set of any number of vertices. %

We will give here the abstract definitions of the most popular graph models used by NoSQL stores in practice: the labeled property graph (LPG) and the resource description framework (RDF). %
We follow \cite[§2.1]{besta2019demystifying}, but \cite[§2]{angles2017foundations} gives similar definitions. %

The LPG model is an extension of the simple graph model that adds metadata (labels and properties) to vertices and edges. %
Properties are key-value pairs, while labels are simple scalar values. %
A LPG \(G\) is a tuple %
\begin{displaymath}
  (V, E, L, l_{V}, l_{E}, K, W, p_{V}, p_{E})
\end{displaymath}

where: %

\begin{itemize}
\item \(L\) is the set of labels; %
\item \(l_{V} : V \mapsto{} \mathcal{P}(L)\) and \(l_{E}: E \mapsto \mathcal{P}(L)\) are the labeling functions for vertices and edges respectively (\(\mathcal{P}\) is the powerset function); 
\item  \(K\) is the set of property keys; %
\item \(W\) is the set of property values; %
\item a property \(p\) is a pair \((k, v)\), with \(k \in K\) and \(v \in W\); %
\item \(p_{V} : V \mapsto \mathcal{P}(K\times{}W)\) and \(p_{E} : E \mapsto\mathcal{P}(K\times{}W)\) are the functions that give the set of properties of a given node and edge, respectively. %
\end{itemize}

RDF \footnote{\url{https://www.w3.org/TR/rdf11-concepts/}} is a data model specified by the W3C and is at its core a collection of triples. %
A triple is composed by a subject, a predicate, and an object. %
A subject \(s\) may be an identifier or a blank node (essentially a dummy identifier), while a predicate \(p\) is always an identifier, and an object \(o\) may be an identifier, a blank nodes, or a literal (a value like a string or a number). %
A triple is thus a connection between identifiers or between an identifier and a literal: %
\begin{displaymath}
  (s, p, o) \in (\text{URI} \cup \text{blank}) \times \text{URI} \times (\text{URI} \cup \text{blank} \cup \text{literal}) %
\end{displaymath} %
where `URI' is the set of identifiers, `blank' is the set of blank node identifiers, and `literal' is the set of literal values. 

According to \cite[§2.1]{besta2019demystifying} there is no one standard graph model, but the last two models are the most popular. %
RDF is a well-defined standard although LPG enables more natural data modeling, according to \cite[§4.10.1]{besta2019demystifying}. %
In \cite{Oracle2014}, the authors propose three different transformations from LPG to RDF to demonstrate that RDF is fully capable of representing any property graph. %
Oracle Spatial and Graph were used in the experimental analysis of the paper but the authors indicated that their methodology can be used in any other Triple Store. %
They identified that the main challenge of the transformation is to represent the key/value properties of property graph edges in RDF. %
Some graph databases, such as Neo4j, DEX, TigerGraph and InfiniteGraph, are based on the labeled property graph (LPG) model. %
On the other hand, RDF has been adopted by other graph databases such as Oracle Database Spatial and Graph Option, Openlink Virtuoso and AllegroGraph. %
We will describe Neo4J and AllegroGraph more in-depth in sections \ref{sec:neo4j} and \ref{sec:allegrograph}. %

\subsection{Storage and Representation}
\label{sec:storage-representation} %

With respect to graph storage, systems are generally divided between native and non-native\cite[§2.4]{davoudian2018survey}. %
Native systems boast graph-aware logical models, while non-native systems rely on non-graph data stores (e.g., a document or key-value store in the case of ArangoDB, a RDBMS in the case of AgensGraph). %
See \cite[§2.2]{besta2019demystifying} for a brief overview of how graph data is represented in non-graph data models. %

Native graph stores may be further subdivided into what graph representations they use\todo[disable]{is it true that we can't we say that nonnative dbs use these representations? I think so, but…}. %
Graph data may be laid out in several ways, the most common ones being the AM (\textit{adjacency matrix}), CSR (\textit{compressed sparse row}) format (commonly used for sparse graphs), AL (\textit{adjacency lists}), and EL(\textit{edge lists}). %
See \cite[§2.4]{davoudian2018survey} and \cite[§2.2]{besta2019demystifying} for a full overview of these representations\todo[disable]{add table with summary of their characteristics?}. %

\subsection{Graph Query Patterns and Languages}
\label{sec:graph-query} %

There are several graph query languages in existence; a non-exhaustive list includes \sparql{}, Cypher, Gremlin, GSQL. %
At the moment, only \sparql{} is standardized by a major institution. %
\footnote{i.e., one of ISO, ANSI, W3C, ECMA, or the IETF.} %

SPARQL 1.1 \footnote{\url{https://www.w3.org/TR/sparql11-query/}} is a declarative language (a SQL-like style), so users can focus on specification and the query engine is responsible for optimal execution. %
The query language is primarily intended for pattern (subgraph) matching rather than path traversal and it is possible to bind variables to generate pattern solutions. %
Property paths \footnote{\url{https://www.w3.org/TR/sparql11-property-paths//}} adds the ability to match connectivity of two nodes by an arbitrary length path with some limitations. %

Cypher has been standardized as openCypher\cite{openCypher}, and ISO is currently working on a graph query language standard%
\footnote{See \url{https://web.archive.org/web/20200613042006/https://www.gqlstandards.org/} for the new standard's webpage and \url{https://web.archive.org/save/https://www.iso.org/standard/76120.html} for its page at the ISO website.} %

Gremlin is a graph traversal query language for property graphs. %
It was designed, developed, and distributed by the Apache TinkerPop project \footnote{\url{http://tinkerpop.apache.org/}}, an open source graph computing framework distributed under the Apache version 2.0 license. %

\cite{angles2017foundations} gives a thorough overview of graph query languages and their fundamentals, while \cite[§3.5, §3.7,§5.1]{besta2019demystifying} gives a more concise summary. %
Regardless of the concrete graph query language in question, all graph queries are built upon two elemental operations: pattern matching \cite[§3]{angles2017foundations} and navigation \cite[§4]{angles2017foundations}. %

Graph pattern matching queries are graph models whose constants may be replaced by variables; the results are the subgraphs of the queried graph that satisfy the constraints imposed by the query. %
As an example, let us say we have a graph \(G = (V, E)\) whose vertices are students and classes, and whose edges represent that a student attends a certain class. %
If we want to now which classes a certain student attends, we issue a query which represents a subgraph \(G' = (V', E')\) of \(G\) where \(V' = {\text{Student}, x}\) and \(E' = {(\text{Student}, x)}\), and \(\text{Student}\) is the identifier of the student in question and \(x\) is a variable. %
The query engine then finds all subgraphs that match the subgraph specified by the query. %

The process of receiving a query and returning its matching results is known as \textit{query evaluation}. %
This definition of evaluation as finding matching subgraphs is informal, since there are several possible ways of deciding what is a match and what is not. %
One common difference among different query evaluation semantics pertain to the possibility of different variables matching to the same entity; some query languages specify that this is allowed, while others restrict different variables to match different entities. %
We do not go into the details of the possible semantics for query evaluations, as this discussion is made by \cite[§3]{angles2017foundations}. %
\todo[disable]{Maybe we should specify some of the possible different semantics, so that we can talk about how cypher and sparql fit into them…} %

Graph navigation is done by so-called path-queries. %
\cite[§4]{angles2017foundations} define a path query \(P = x \to^{\alpha} y\), where \(x\) and \(y\) are the start and end vertices (or variables), and \(\alpha\) are the restrictions on the path from \(x\) to \(y\). %
The simplest path query places no specific restrictions on the path other than its existence, but more complex restrictions are possible. %
More complex path queries usually are specified by some variant of Regular Path Queries (RPQ), which are a regular expression language for path queries. %

\todo[disable]{talk about path query evaluation? what will we say for Neo4j? (I'm sure there are resources for sparql, but for cypher I'm not so sure…)} %
\todo[disable]{semantics of path query evaluation}

Although pattern matching and path finding are the two hallmarks of graph queries, languages differ widely on their support for them. %
Some languages like \sparql{} are more oriented towards matching, while others like Gremlin are more focused on path finding. %


\subsection{Graph Database Technology Dimensions}
\label{sec:graph-db-dimensions} %

\begin{figure}
\centering
\includegraphics[width=0.6\textwidth]{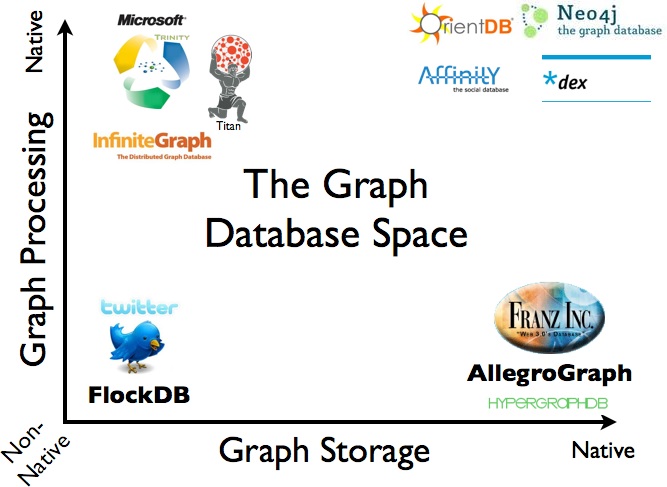}
\caption{The Graph Database Space}
\label{fig:graphdbspace}
\end{figure}

Two dimensions of graph databases are worth to observe when analyzing graph database technologies: the underlying storage and the processing engine (see Figure \ref{fig:graphdbspace}). %

As explained in \myref{sec:storage-representation}, graph storage may either be native or non-native. %
Native storage is optimized and designed for graph data, while non-native storage may serialize the graph data into a relational, object-oriented, or other types of NoSQL stores. %

Processing engines can also be native or non-native. %
The native ones leverage the index-free adjacency property to traverse through the graph efficiently, as nodes connected by their relationships physically ``point'' to each other in the database. %
The creators of Neo4j define `index-free adjancency' thus:
\begin{displayquote}[\cite{robinson2015graph}]
  A database engine that utilizes index-free adjacency is one in which
  each node maintains direct references to its adjacent nodes.
\end{displayquote} %

Triple stores, do fall under the general category of graph databases, but aren't native processing because they do not support index-free adjacency. %
Such graph databases store triples as independent artifacts. %
This characteristic enables horizontal scaling but limits their efficiency in relationship traversal. %
To perform traversals, triple stores tend to create connected auxiliary structures, which add latency to query processing. %
According to \cite{graphbook}, this particularity suggests that Triple stores are more suitable for Analytics scenarios, where latency is a secondary consideration, rather than Online Transaction Processing (OLTP). %

\subsection{Consistency levels}
\label{sec:consistency}

When we talk about databases there are two meanings to consistency. %
Consistency as the \texttt{C} in \textit{ACID} acronym is a property of transactions, and is the preservation of database consistency by the successful execution of the transaction. %
This means that if a transaction finds a consistent database at the beginning of its execution, it leaves out a consistent database at the end of its execution. %
In this case consistency is mainly application-defined (things like ``all students must have an enrollment identifier''), but also includes things like the preservation of foreign-key constraints. %
The consistency in \textit{ACID} is thus mostly the responsibility of programmers who write the software that interacts with the database, and the DBMS module that enforces integrity constraints\cite[§20.3]{elmasri2016fundamentals} (if they are enforced at all, which is not always the case in NoSQL stores). %

The other meaning of consistency which is common in the database research is that of the \textit{C} in the \textit{CAP} theorem. %
In distributed systems where data is replicated across several nodes the possibility of different nodes storing different copies of the same data item is real. %
Consistency in this case is the property of such a system appearing to be have a single copy of each data item, so interactions with it will not receive outdated or inconsistent values\cite[§24.2]{elmasri2016fundamentals}. %

Because the former consistency is mostly the responsibility of end-users while the latter consistency is a responsibility of the DBMS, the consistency alluded to by the \textit{CAP} theorem is the one that concerns us here. %
Just as is the case with the property of isolation (see \myref{sec:isolation}), consistency is not an absolute concept; it admits different levels, of which \textsc{strict consistency} (in theory) and \textsc{linearizability} or \textsc{atomic consistency} (in practice) are the highest levels, and \textsc{eventual consistency} is the lowest. %

\cite{viotti-consistency} defines \textsc{linearizability} informally as a consistency level that guarantees that ``each operation shall appear to be applied instantaneously at a certain point in time between its invocation and its response.'' %
Refer to \cite{viotti-consistency} for formal and informal definitions of different consistency guarantees. %
This allows operations that do not overlap their start and acknowledgement times to be ordered unambiguously, and is the consistency level traditionally offered by RDBMSs. %
NoSQL systems on the other hand need greater availability and thus sometimes sacrifice some consistency to achieve better availability. %
\cite{diogo2019consistency} surveys the consistency level of several NoSQL systems. %

\subsection{Isolation guarantees}
\label{sec:isolation} %

The topic of isolation is not specific to graph databases, but to any transactional system, including RDBMSs and NoSQL DBMSs. %
In this section we define what transactional isolation is and give an overview of the possible guarantees a system can offer about it. %

\cite{elmasri2016fundamentals} defines transactional isolation as the property that makes ``A transaction […] appear as though it is being executed in isolation from other transactions, even though many transactions are executing concurrently.'' %

Transactional isolation is not an absolute concept: it can be achieved at different levels. %
These describe how transactions may (or more often, how they may not) interfere with each other. %
\cite{berenson1995critique} defines several levels of isolation guarantees in terms of the kinds of schedules they prohibit, all the while criticizing the levels defined by the ANSI SQL standard. %
\cite{crooks2017seeing} provides other definitions that are client-centric: they depend only on states that are observable by the users of the systems. %
Such definitions do not `leak' implementation details and are arguably more easily understood; it also helped \cite[§5.2]{crooks2017seeing} to prove that several variants of snapshot isolation were actually offering the same guarantees to system end-users. %

Serializability is the greatest possible isolation level, and corresponds to absolute isolation (i.e., as if one is executed after the other, and not at the same time). %
We will discuss the isolation level of Neo4j in \myref{sec:neo4j-isolation}. %

\section{Graph Database Systems}
\label{sec:graph-database-systems}

There are works in the literature dedicated to NoSQL systems, some focusing on listing and comparing NoSQL databases features, exposing their advantages and disadvantages\cite{graphcomp2017, persisting2017}. %
There also exist several surveys, some of them included graph models as NoSQL databases \cite{davoudian2018survey} and others which did not \cite{SurveyWithoutGraphs}. %
Some studies are dedicated to the creation of a taxonomy of different databases \cite{besta2019demystifying} or to define and analyze graph database models \cite{angles2008survey, renzo2012comparison}. %
In section \ref{sec:graphdboverview} we will overview a few of the studies on graph databases. %

There are some NoSQL Graph Database Systems frequently mentioned in those works. %
In the section \ref{sec:graphdbexamples} we will give an overview of few of them and in section \ref{sec:implementations} we will analyse in detail how two different graph database management systems\emdash{}AllegroGraph and Neo4j\emdash{}fit into the taxonomy described in this section. %

\subsection{Graph Database Academic Overview}
\label{sec:graphdboverview} %

Graph database models were compared with other database models (Network, Relational, Semantic, Object-Oriented, Semi structured) by \cite{angles2008survey} to distinguish their applications and characteristics (abstraction level, base data structure, and information focus). %
The relational database model introduced a separation between the physical and logical levels (abstraction levels), gave a mathematical foundation to the data modeling discipline and introduced a standard query and transformation language, SQL. %
But its data structure (relations) has to be known in advance and its fixed schema makes for difficult extensions or remodelings. %
The SQL language was not designed to explore the underlying graph of relationships among the data, such as paths, neighborhoods, patterns. %
This makes querying graph data stored in a regular RDBMS expensive because of complex join operations and index lookups\cite[§2.4, footnote 36]{davoudian2018survey}; this contrasts with how easy it is to embed graphs in the relational model. %

In user abstraction level, database designers using semantic database models like entity-relationship model can represent objects and their relations naturally using conceptual abstractions such as aggregation, instantiation, and hierarchies. %
These types of models, whose information focus is in schemas plus relations, were important for graph model research, whose information focus is in the data plus relations, %
\todo[disable]{I'd rephrase this; semantic databases xx aqui é modelo de banco de dados e não sistema banco de dados, modelo como o EER é semantico xx are not the only ones whose focus is on data relationships (plus, isn't it weird to say that a database focuses on data?)} %
since it reasons about the underlying graph structure presented in the relationships among the modeled entities. %
Semi structured models like XML have an ordered-tree-like structure, which is a restricted type of graph. %
The authors also compared graph database model proposals found in the literature, from 1984 until 2002, using a theoretical perspective in terms of modeling aspects: data structures, query languages and integrity constraints. %
A common issue among the proposals is the level of separation between schema and instances and, in most cases, they are clearly distinguished. %

An evaluation of the graph models with a more practical perspective can be found in \cite{renzo2012comparison}. %
In this study, the authors distinguish graph stores from graph databases. %
For them, graph stores are implementations that provide basic facilities for storing and querying graphs whilst graph database must offer advanced features like a query optimizer and backups, as database management systems do. %
\cite{renzo2012comparison} compares six NoSQL graph databases, AllegroGraph, DEX, HypergraphDB, InfiniteGraph, Neo4J and Sones, and three graph stores, Filament, G-Store and vertexDB, concentrating on their data model features in terms of the same modeling aspects. %
Compared with mature databases, like relational databases that uses SQL, the authors observed that there is a lack of query language standardization for graph databases and the main development focus is to provide APIs for popular programming languages. %

He also stated that integrity constraints, like type checking, are poorly studied in graph databases and, although the support for schema evolution is a common justification for this absence, data consistency in a database might be equal to or even more important than flexibility. %
Besides, this article defined a set of essential graph queries: adjacency (Node/edge adjacency and k-neighborhood), reachability (Fixed-length paths, Regular simple paths and, Shortest path), pattern matching and summarization. %
Analyzed the support of the selected NoSQL databases to answer such queries, he observed that none of the systems provides k-neighborhood nor summarization out-of-the-box. %

An approach to NoSQL graph databases design given its customary schemalessness is proposed in \cite{akoka-four-vs-design}. %
They affirm that although schemalessness is considered an advantage of NoSQL databases, query optimization and data integrity are best performed in the presence of schemas. %
Although we are inclined to agree with this statement, we found \cite{akoka-four-vs-design}'s argumentation lacking, nor did they seem to provide references to support this view. %
Their methodology takes into account the four V's of big data: volume (data quantity), variety (many different types of data), velocity (the speed that data is made available and accessible%
\footnote{\textbf{NB}: not the speed at which it is produced}%
), and veracity (different levels of data inconsistency and incompleteness). %

The forward engineering approach proposed in \cite{akoka-four-vs-design} is based on model-driven architecture (MDA) to guide the development of conceptual, logical, and physical models with the support of a set of transformations rules, to move from one level to another, and two meta-models: V’s EER conceptual meta-model and EER logical property graph meta-model. %
The logical model generated by their method propagates the four V's data inputted by the user, so that the physical models can be generated according to their specification. %
This allows the creation of dummy databases that follow the data model and that have realistic 4 V's parameters, which can be very useful to benchmark a data store for a prospective use case. %

\cite{persisting2017} compares several NoSQL databases across 7 dimensions (Persistence, Replication, Sharding, Consistency, API, Query Method, and Implementation Language). %
The comparison aims to assist in the choice of which one is the most appropriate for a given set of requirements. %
In the Graph-oriented databases group, the selected databases were: Neo4J, Infinite-Graph, InfoGrid, HyperGraph, AllegroGraph, and BigData. %

The authors mention that Neo4J, instead of Allegro Graph, uses \sparql{} for querying data stored in RDF format and Gremlin to graph traversal over graphs stored in various formats\cite[§7]{persisting2017}; this is factually wrong: Neo4j uses the Cypher query language for declarative queries, and offers a Java API for imperative graph manipulation. %
With respect to the CAP theorem, Neo4J and Allegro Graph were considered to compromise some consistency in exchange for AP (Availability-Partition Tolerance). %
Neo4j supports limited causal consistency (see \myref{sec:neo4j-consistency}), and both stores support eventual consistency. %

\cite{graphcomp2017} analyses five NoSQL databases, of which two are multi-model (OrientDB and ArangoDB) and three are graph models (AllegroGraph, InfiniteGraph and Neo4J). %
They compared graph databases across what they take to be the most important features of a graph database: the flexibility of their schemas (the ability to add new type of vertices and edges without impact in the previously stored data)\todo[disable]{it would be nice to reference here where we define this}, the query language they use (which one suits best the needs of querying graph data)\todo[disable]{any query language?}, their sharding support, how they backup their data, their support of multiple data models, how well they scale and how easy it is to deploy them to the cloud. %
Supporting multiple data models was defined by the authors as the capability of the graph database to store unstructured data, and to visualize data relationships in the form of graphs, key-value pairs, documents or tables. %
Each NoSQL feature was graded in a scale from 0 to 4, for a total of 36 points, based on the authors' experience (subjective analysis) and a literature review. %
By this analysis, they concluded that Neo4J and ArangoDB stand out for their functionalities, both achieving a 26 points score. %
However, they recommend Neo4J over ArangoDB because of its enhanced implemented features although it did not support sharding at the time. %
\footnote{Neo4j supports (manual) sharding since version 4, see \url{https://neo4j.com/release-notes/neo4j-4-0-0/}} %

In \cite{besta2019demystifying} the authors broadened the study to any database system that enables storing and processing graphs, in other words, graph database systems. %
They included native graph databases and other types of databases with different data models that do not target specifically graphs but are used in various systems to model and store graphs. %
Some systems are simple graph frontends to other kinds of storage management systems (non-native), while others are custom-built for the purpose of storing and managing graphs (native). %
Implementations also differ in what kinds of graphs they work with. %
\todo[disable,author=bruno]{should probably refer to what we've written already or scrap this part, since it's repetitive} %
While graph theory gives us different models for graphs, which can be undirected, directed, labeled, attributed, multigraphs, and hypergraphs (among others), engineering practice gives us two common graph models: RDF and property graphs. %
Graph stores may also differ in how they organize the data they hold, specially in which kind of graph representation they use (usually a variation of adjacency matrices, adjacency lists, or edge lists). %
Other differences lie in the query language used (there is no standard language for graph querying), in how queries may be executed (can they be run concurrently? Can they be parallelized?), how data is partitioned (or not) among nodes, and in how transactions are supported (how ACID-compliant they are? Are they run online or offline?) %

The selected 45 graph database systems were grouped in 9 categories (RDF store/Triple Store, Tuple store, Document store, Key-value store, Wide-column store, RDBMS, OODBMS, Native graph store/databases, and Data hub) based on their general database engine and analyzed in 5 key dimensions (data model, data organization, data distribution, query execution, and type of transactions). %
AllegroGraph was classified as a RDF Store/Triple Store, since the main data model used is RDF triples. %
Neo4j was classified as Native Graph Database, since LPG (Labeled Property Graph) is the main data model used.\todo[disable,author=bruno:]{the model could be LPG and the database could not be native, these things are orthogonal} %
The authors used a different definition for graph databases/stores: \textit{systems that were specifically build to maintain and process graphs}. %
In this survey, OrientDB and ArangoDB were classified as Document Stores with Multi-Model support (LPG and RDF) and Oracle Spatial and Graph was classified as Relational Database Management Systems (RDBMS) that also has Multi-Model support using a row-oriented storage. %

\subsection{Graph Database Examples}
\label{sec:graphdbexamples} %

\todo[disable]{contrast these with neo4j and allegro} %

FlockDB is an open source graph database developed by Twitter based on MySQL. %
It has horizontal scaling capabilities (distributed and fault-tolerant) for managing wide but shallow network graphs \cite{besta2019demystifying}, i.e., social graphs representing who is following whom and who is blocking whom (their adjacency list) and answer simple graph queries, such as finding mutual friends. %
The latest release is 1.8.5 from March 2012. %
FlockDB is not a general-purpose database like Neo4j and AllegroGraph. %
As stated in its description\footnote{See \url{https://github.com/twitter-archive/flockdb} (accessed on 2020-06-30) for more information.} %
FlockDB's graph model is that of a directed graph, and may not have two edges between the same nodes. %
An edge contains only two metadata: a state (one of normal, removed, or archived) and a position (usually a timestamp, which is used for sorting). %
Because of its simpler use-case, FlockDB can scale better; in 2010 Twitter was using a FlockDB cluster to store more than 13 billion edges, and this cluter withstood a peak traffic of 20k writes/second and 100k reads/second. %

Microsoft Trinity was launched in 2013 and its lastest release is from November 2017. %
It has since been rebranded as Microsoft GraphEngine.\footnote{See \url{https://www.microsoft.com/en-us/research/project/trinity} and \url{https://www.graphengine.io/}, accessed on 2020-06-30.}. %
It uses hypergraph as the data model \cite{graphsurvey2015}. %
As a non-native graph store, graph data is represented through adjacency lists stored in key-value stores. %
Consistent hashing is used to map a key to one machine of the shared distributed memory cloud, globally addressable, across multiple servers (cluster). %
Data persistence is supported by Trinity File System (TFS), a distributed shared file system like HDFS\cite{davoudian2018survey}. %
GraphEngine uses its own query language, the language integrated knowledge query (LIKQ). %

TitanDB is an open source project of a graph database system based on property graph model. %
Hadoop framework is used in the backend of graph processing and Gremlin (a graph traversal query language) is used for graph retrieval and manipulation \cite{graphsurvey2015}. %
TitanDB also enables other storage backends like Apache Cassandra. %
JanusGraph is the continuation of the Titan project led by the Linux Foundation \cite{besta2019demystifying}. %

Infinite Graph is a cloud enabled graph database as it distributes a single graph database across multiple machines \cite{graphsurvey2015}. %
The property graph model supports integrity constraints: types checking and node/edge identity and the definition of node and relation types at the schema level \cite{renzo2012comparison}. %
Development on Infinite Graph has since been discontinued, and has been spread across two other products of the same company, ThingSpan and Objectivity/DB.%
\footnote{See \url{https://www.objectivity.com/products/infinitegraph/}, accessed on 2020-06-30.} %

DEX allows the management of persistent and temporary graphs. %
Bitmaps and other secondary index structures are used to enhance the performance of large graph structures. %
It supports integrity constraints: type checking, referential integrity and node/edge identity and the definition of node and relation types at the schema level \cite{renzo2012comparison}. %
Its graph data model can be called “Labeled and directed attributed multigraph” because it enables directed or undirected edges and more than one edge between two nodes \cite{graphsurvey2015}. %
DEX was renamed as Sparksee in 2014. %
Sparksee claims to support ACID transactions with a \textsc{SERIALIZABLE} isolation level.\footnote{See \url{http://www.sparsity-technologies.com/\#sparksee} (accessed on 2020-06-30).} %
Sparksee has no graph query language, depending on custom APIs to obtain and modify information from the graph.%
\footnote{See the Sparksee user manual: \url{http://sparsity-technologies.com/UserManual/API.html} (accessed on 2020-06-30).} %

\section{Implementations}
\label{sec:implementations} %

In this section we delve into some of the core characteristics of two popular but very different graph databases: Neo4j and AllegroGraph. %

\subsection{Neo4j}
\label{sec:neo4j} %

Neo4j is native graph database first released in 2007 and written in the Java programming language. %
It is the most popular graph database as of June 2020, according to the DB-engines ranking.\footnote{See \url{https://db-engines.com/en/ranking/graph+dbms} (accessed on 2020-06-30).} %
In this section we detail how Neo4j represents and stores graphs (\myref{sec:neo4j-storage-representation}), the query language support it provides (\myref{sec:neo4j-query}), and what kinds of consistency (\myref{sec:neo4j-consistency}) and isolation (\myref{sec:neo4j-isolation}) guarantees it offers. %

\subsection{Storage and Representation}
\label{sec:neo4j-storage-representation} %


Neo4j is native graph store that represents graphs using an adaptation of the AL (adjacency list) and EL (edge list) formats. %
(See \myref{sec:storage-representation} for a recapitulation of these concepts.) %
Neo4j stores graphs in disk by distributing them in several record files (one for nodes, one for relationships, one for properties, etc.)~composed of fixed-size records\cite[Chapter 6]{robinson2015graph}.%
\footnote{This description of Neo4j in-disk storage is valid for its second version, since this is the one that we could find bibliographic references for. %
  By inspecting the release notes of Neo4j versions 3 and 4 we found no evidence of significant deviations from this description.} %

We affirm that Neo4j employs an adaptation of the AL and EL representation formats because the actual representation depends on how you look at these record files. %
By considering solely the record file for relationships we would have the impression that the representation format is that of an edge list, since this record file is essentially a list of relationships. %
By considering the record file for nodes on its own we would conclude that the graph representation format used by Neo4j is that of an adjacency list, since this file is essentially a list of nodes, each (among other things) storing a pointer to its list of relationships. %

This representation used by Neo4j offers advantages similar to those of the AL and EL formats: node and edge insertion is done in constant time, and the space used is linear on the size of the graph. %
An additional advantage is that nodes and edges can be looked up by their identifiers in constant time (this is only true of nodes in the AL format). %
The Neo4j representation also inherits a pitfall of these representations: checking if a node is a neighbor of another is done in linear time on the number of relationships of the node, which in the worst-case is the number of relationships of the whole graph. %

The advantages of the representation format used by Neo4j come mainly from two techniques: letting (node, relationship, etc.)~identifiers be the index of their respective record in the store file, and making abundant use of pointers. %
We detail these techniques in the next paragraphs. %

Letting node and relationship identifiers be mere indices of their respective records allows us constant time access to them. %
As an example, if a node references a relationship with ID 3, we know it is the third record on the relationship store file. %
Because records in this file have a fixed size, we can compute the byte offset to where the record starts and thus access it in constant time. %
This characteristic is what affords Neo4j the so-called `index-free adjacency property' defined in \myref{sec:graph-db-dimensions}. %

The simplicity of looking up a node in a database that has the index-free adjacency property contrasts with the complexity of doing the same operation in a database that doesn't: we would have to look up the ID in an index, which is usually done in logarithmic time. %

\begin{figure}\centering
  \includegraphics[width=\textwidth]{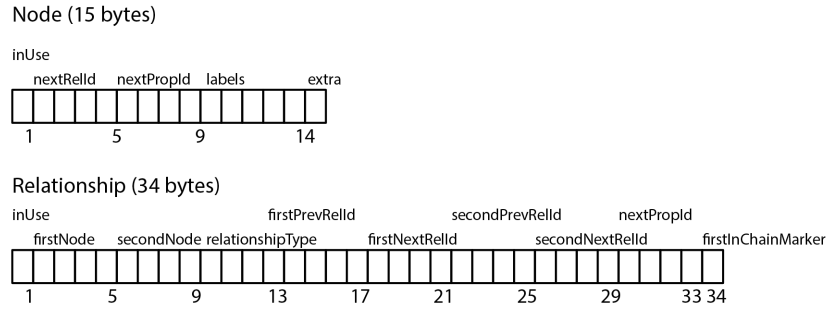}
  \caption{Binary structure of node and relationship store files
    in Neo4j versions 2.* \cite{robinson2015graph}}
  \label{fig:record-structure}
\end{figure}

\begin{figure}\centering
  \includegraphics[width=\textwidth]{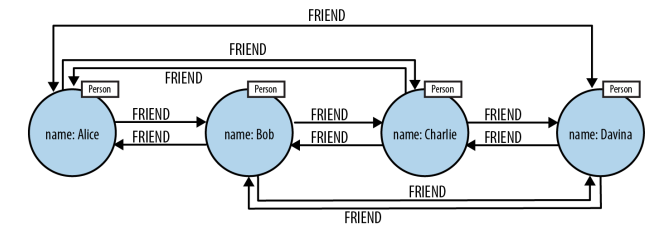}
  \caption{Graph representation in Neo4j \cite{robinson2015graph}}
  \label{fig:relationships-in-neo4j}
\end{figure}

\Cref{fig:record-structure} shows the binary structure of a node and a relationship record in Neo4j (versions 2.*). %
As we can see, a node record has a pointer to the identifier of its first relationship (and to its first property too). %
Along with the fact that relationships have pointers to the next and previous relationships for both its start and end nodes, what we have in practice in a node record is a pointer to a doubly-linked list for its relationships (see \cref{fig:relationships-in-neo4j} for a schematic example). %
We do not consider node/relationship properties here, but \cite{raj2015neo4j} gives an overview of their on-disk layout. %

\subsection{Query language and support}
\label{sec:neo4j-query} %

Neo4j uses the Cypher query language, which has been standardized under the name `openCypher'. %
We did not find any document stating differences between openCypher and the Cypher implementation of Neo4j, if there any. %
\cite{francis2018cypher} is the most complete reference paper on Cypher, including a semantic model of the language and bibliographic references to its implementation details. %

Cypher is a declarative query language, and employs pattern matching heavily using a DSL inspired by ascii art. %
The example query we provided in \myref{sec:graph-query} is expressed in Cypher as in \myref{fig:cypher-example}. %
Note that \texttt{:Student} and \texttt{:Class} are node lables, \texttt{:Attends} is a relationship label, and \texttt{\textdollar{}studentId} is a parameter specifying the identifier of the student in the example. %
The result is returned as rows containing a field named `class', but could also be modified to return one row with classes aggregated into a list. %

\begin{figure}\centering
\begin{verbatim}
MATCH (s:Student)-[:Attends]->(c:Class)
WHERE id(s) = $studentId
RETURN c AS class
\end{verbatim} %
  \caption{Cypher query returning the classes a student attends}
  \label{fig:cypher-example}
\end{figure} %

Cypher is also clause-based, having a resemblance to SQL. %
An extension library\cite{APOC} allows the language to be extended using the Java programming language or Cypher itself to define new functions and procedures, which can then be called from Cypher using the \texttt{CALL} clause. %
This is equivalent to the stored procedures and user-defined functions common in some RDBMSs. %

\subsection{Consistency guarantees} %
\label{sec:neo4j-consistency} %

\begin{figure}
  \centering
  \includegraphics[width=0.65\textwidth]{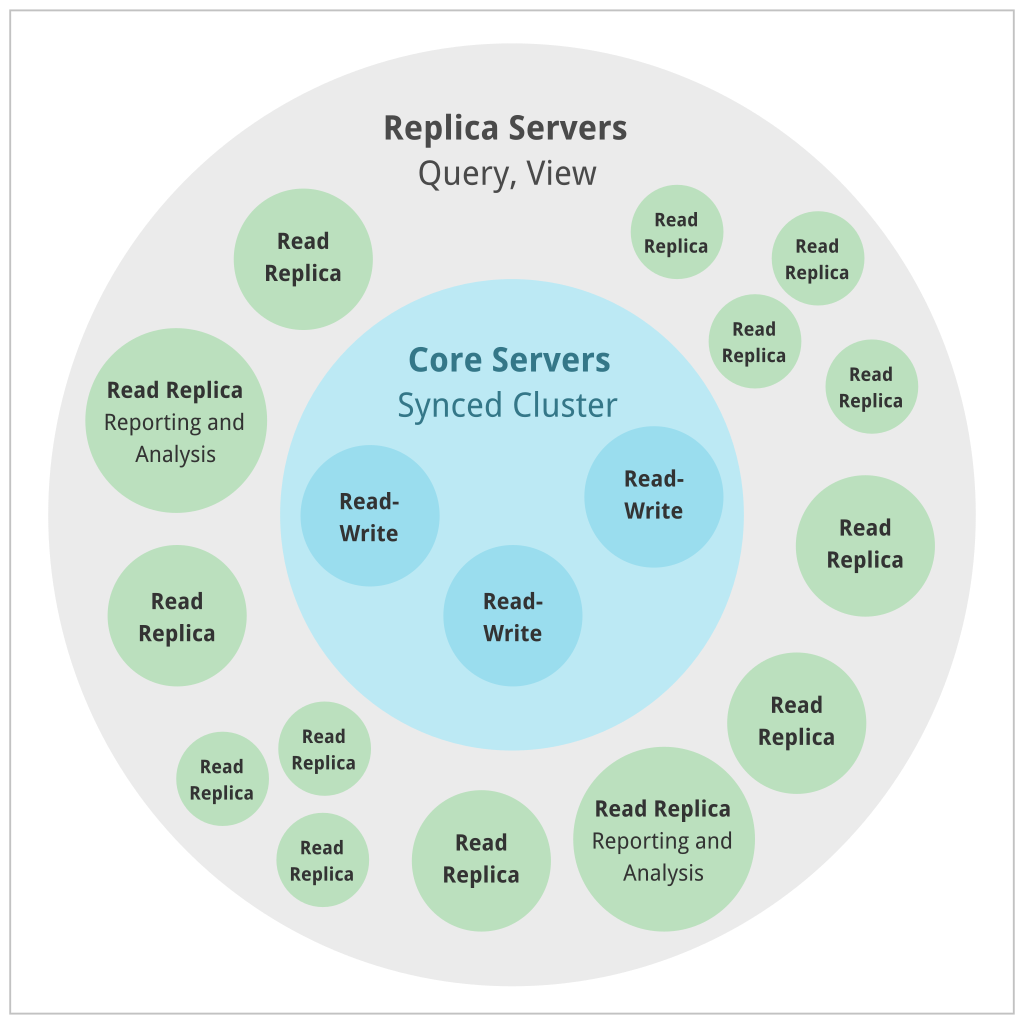}
  \caption{Typical Neo4j cluster architecture \cite[§7.1]{neo4j-operations-manual}}
  \label{fig:neo4j-cluster}
\end{figure}

To talk about consistency as alluded to in the CAP theorem, we must first understand how Neo4j works in cluster mode. %
Neo4j clusters are composed of two types of nodes: core and replica (see \myref{fig:neo4j-cluster}). %
Core nodes are the only ones capable of performing writes to the transaction log, while replica nodes (usually more numerous) are in charge of answering read-only queries.\footnote{This clustering architecture assumes that most of the query workload is read-only.} %
Core nodes replicate and synchronize transactions amongst themselves using the Raft protocol\cite[§7, appendix C]{neo4j-operations-manual}. %
The Raft protocol is a consensus algorithm designed to be easily understood and implemented\cite{ongaro2014search}, and was chosen by Neo4j for implementation for this very reason\footnote{See `Causal consistency for large Neo4j clusters', a presentation by Neo4j's Jim Webber at \url{https://www.youtube.com/watch?v=Vcl9Vq0XoUY} (accessed on 2020-06-22)}. %
Neo4j clusters can tolerate faults up to a minority of its core nodes, a characteristic derived from the Raft protocol. %
Despite providing fault-tolerance, using a consensus algorithm also introduces overhead in communications, which is why the number of core nodes in a cluster is usually small. %
Updates to the graph are shipped from the core nodes to the replicas faster, because there is no consensus algorithm involved; the replicas receive the transactions that were included in the transaction log by the core nodes and reproduce their operations on the replica's copy of the data. %

When operating as a cluster Neo4j offers the \textsc{causal} consistency level\cite[§7.1.3]{neo4j-operations-manual}. %
In this consistency model, an ordering between causally-related operations is imposed for all nodes\cite{viotti-consistency}. %
\Myref{fig:causal-inconsistency-example} shows a transaction history from three different clients (C1, C2, and C3) that shows an instance of causal inconsistency. %
In this example, causal consistency is violated because C2 has written to \texttt{y} after having read from \texttt{x}, C3 may not read a stale value of \texttt{x} if it has already read the up-to-date value of \texttt{y}. %

In the case of Neo4j, causal consistency is implemented using bookmarks: transactions acknowledged by the cluster return a bookmark; the bookmark can be sent along with subsequent transactions to establish their causal relatedness; a node that has not seen the bookmark before (and thus not processed its transaction) may not process the current transaction\cite[§7.1.3]{neo4j-operations-manual}. %
The use of bookmarks may seem to create a cognitive overhead for the application programmer, but its use is in fact invisible for the most part, as bookmarks are set and propagated by the driver APIs automatically\cite[§3.5]{neo4j-driver-manual}. %

\begin{figure}\centering
  \begin{tabular}{ccccccc} %
    \textbf{C1}: & R: x=x$_{0}$ & W: x=x$_{1}$ &         &         &         &\\\cline{2-7}
    \textbf{C2}: &         &         & R: x=x$_{1}$ & W: y=y$_{1}$ &         &\\\cline{2-7} %
    \textbf{C3}: &         &         &         &         & R: y=y$_{1}$ & R: x=x$_{0}$\\%
  \end{tabular}
\caption{Transactions from two database clients that exemplify causal inconsistency}
\label{fig:causal-inconsistency-example}
\end{figure} %

In the absence of bookmarks, Neo4j degenerates to the \textsc{eventual} consistency level. %
At this consistency level, cluster nodes are guaranteed to converge towards the same database state (in the absence of updates). %
The time taken for this convergence is not specified by the definition, and in the case of Neo4j we did not find any statement or data on this topic. %

\subsection{Isolation guarantees}
\label{sec:neo4j-isolation} %

Version 4 of Neo4j offers only one possible isolation level, that of \textsc{read committed}\cite[§3.3. Isolation levels]{neo4j-java-reference}. %
See \myref{sec:isolation} for a definition of transactional isolation. %
In this section we use definitions, notation, and examples from \cite{berenson1995critique}. %

The \textsc{read committed} isolation level prohibits schedules that perform dirty writes and dirty reads\cite{berenson1995critique}. %
To describe what kinds of schedules these are, we will schematize a schedule \texttt{S} to be a list of operations, where operations are represented by a letter (\texttt{w} for write, \texttt{r} for read, \texttt{a} for abort, and \texttt{c} for commit operations), an integer identifier of the transaction the operation belongs to, and (optionally) the data item being operated upon between square brackets. %
In this notation ``…'' stands for any number of operations that do not concern us. %

A schedule performs a dirty write if it contains a transaction that modifies a data item that is subsequently modified by another transaction before the original transaction terminates. %
Using the schedule notation described above, any schedule matching the schema in \myref{fig:dirty-write-schedule} is performing a dirty write and is guaranteed to be disallowed by a \textsc{read committed} isolation level\cite{berenson1995critique}. %

As noted by \cite{berenson1995critique} and exposed in \myref{fig:dirty-write-schedule}, the original transaction does not have to abort for problems to arise, for database consistency requirements might be broken even when both transactions commit successfully. %
If a database has a consistency constraint where the value of \texttt{x} must equal that of \texttt{y}, this invariant might be broken by a schedule such as the one in \myref{fig:dirty-write-subtle-example}. %

A schedule performs a dirty read when they have a transaction modifying a data item that is subsequently read by another transaction before the original transaction has the time to terminate (by either aborting or committing). %
This English language description of a dirty-reading schedule can be written using the schedule notation described above as in \myref{fig:dirty-read-schedule}. %
An example transaction performing a dirty read can be seen in \myref{fig:dirty-read-subtle-example}. %
If a database has a consistency requirement that the sum of \texttt{x} and \texttt{y} must equal a constant value, the schedule seen in \myref{fig:dirty-read-subtle-example} will break such an invariant, since the sum of \texttt{x} and \texttt{y} seen by the two transactions is different. %

\begin{figure}\centering
\schemadule{w1[x]…w2[x]…((c1 or a1) and (c2 or a2) in any order)} %
\caption{Schema for schedules that perform dirty writes}
\label{fig:dirty-write-schedule}
\end{figure} %

\begin{figure}\centering
\schemadule{w1[x] w2[x] w2[y] c2 w1[y] c1} %
\caption{Example schedule were a dirty write is performed}
\label{fig:dirty-write-subtle-example}
\end{figure} %

\begin{figure}\centering
\schemadule{w1[x]…r2[x]…((c1 or a1) and (c2 or a2) in any order)} %
\caption{Schema for schedules that perform dirty reads}
\label{fig:dirty-read-schedule}
\end{figure} %

\begin{figure}
  \centering
  \schemadule{r1[x=50] w1[x=10] r2[x=10] r2[y=50] c2 r1[y=50] w1[y=90] c1}
  \caption{Example schedule were a dirty read is performed}
  \label{fig:dirty-read-subtle-example}
\end{figure}

Despite the default transactional isolation level of \textsc{read committed} offered by Neo4j, greater levels of transactional isolation may be implemented manually. %
This is done by employing locks to guarantee a greater level of isolation\emdash{}we know already that using a protocol like 2-phase locking guarantees serializability, the greatest level of isolation. %

There are two ways to set locks in Neo4j\cite[§3.3. Isolation levels]{neo4j-java-reference}: one option is to use the Java API, which allows manual setting of locks, and the other is to use the Cypher query language. %
Cypher creates locks automatically in simple cases, and an expert user can manually set properties in the graph to provoke this automatic locking; this only works for write locks, however. %
Besides this restriction, setting locks is not explicitly available in the Cypher query language, and the use of locks is an implementation detail of Neo4j, so this does not carry over to other stores that use Cypher, following the openCypher\cite{openCypher} standard. %

\subsection{AllegroGraph}
\label{sec:allegrograph}

AllegroGraph is a multi-model NoSQL database: document in JSON / JSON-LD (starting from version 6.5) and graph in RDF / OWL. %
The first known version of AllegroGraph date from the end of 2004. %
Version 4.14 was released on July 2014, 6.5.0 on March 4th, 2019, and the last one stable version 7.0.1 is from June 8th, 2020. %
It complies with W3C standards for the Semantic Web and this characteristic can be seen as an advantage since data migration to other Triple Stores that adopted the same standards is facilitated. %
AllegroGraph Free and Enterprise Editions are available on Amazon Cloud Services (Amazon EC2). %

Graphic data can be loaded in various RDF formats: N-Quads, N-Triples, RDF / XML, Trig, TriX and Turtle as well as CSV files. %
AllegroGraph supports several specialized datatypes for efficient storage, manipulation, and search of Social Network, Geospatial and Temporal information. %
AllegroGraph also supports the Integers, Unsigned Integers, Floating point, Decimals, and Times and Dates types from the XML Schema Built-in Datatypes. %
There is two ways to supports datatypes (1) tagging data with the type but the data is actually stored as a string and (2) encoding the data so it isn't stored as strings but in a special format to allows fast lookup, range and order queries on these encoded datatypes avoiding full scan. %
Appendix \ref{sec:appendixA} provides examples of tagging data with its type carried out in the tests. %
AllegroGraph do not impose the obligation to define any schema before data load and it accepts a variety of RDF formats (plain based and XML based) so it is not clear why AllegroGraph was classified as bad (1 point) the in flexible schema criteria by \cite {graphcomp2017}. %

AllegroGraph implements the ACID properties (atomicity, consistency, isolation, and durability). %
In terms of atomicity, rollback and commit work the same way as RDBMS. %
AllegroGraph don't implement integrity constraints like RDBMS so the database has no influence on the application-level consistency of the data but it guarantees that every transaction will take the database from one consistent state to another. %
Snapshot isolation model, where each transaction accesses a snapshot of the persistent database state as of the time the transaction begins, guarantees that one transaction cannot affect concurrent others. %
Although, it is important to mention that no triple locking is performed so it cannot guarantee write concurrency. %
AllegroGraph 7.0.1 official documentation says that the database \textit{“guarantees that every commit operation will take the database from one consistent internal state to another consistent internal state”}. %
Then we can conclude that AllegroGraph implements Strong Consistency instead of Eventual Consistency as mentioned by \cite{persisting2017} in table 7. %
Although Strong Consistency is not a common characteristic of NoSQL Databases, considering the databases analyzed \cite{besta2019demystifying} we were able to see that: 7 in 10 Triple Stores are ACID. %
When a commit operation succeeds, the updates made by the transaction are written to the transaction log (Recoverability). %
Log I/O operations are executed periodically (Checkpointing) to promote a permanent effect on the persistent database state. %
To ensure that distributed transactions in AllegroGraph and in external databases will all happen (be commited) or will all fail (be rolled back), AllegroGraph supports two-phase commits (2PC) using a blocking protocol managed by external transaction coordinator. %

The storage is non-native, that is, edges aren't stored in the adjacency list format, but it supports automatic indexing of the triples in addition to indexing free text values in the triples, even allowing the filter of predicates. %
The graphs in the RDF model were stored originally in the form of triples (also called assertions) as specified: subject (s), predicate (p) and object (o).
W3C standard oriented Semantic Graph Databases to store RDF as ‘Quads’ (Named Graph, Subject, Predicate, and Object) although the ‘Triple Store’ terminology is still in use. %
In 2012 AllegroGraph was classified by \cite{renzo2012comparison} without support for Attributed graphs (graphs where nodes and edges can contain attributes for describing their properties. %
Since version 4.14 (2014), AllegroGraph supports named graphs for triples, these type of graph can have attributes stored as triples too. %
Named graph approach was designed to store metadata about triples in a richer model. %
The value of a graph can itself be a node, and the subject of various triples. %
So each triple also has the graph field (g) and an ID (i). %
Appendix \ref{sec:appendixA} provides examples of inserting and retrieving named graphs carried out in the tests.

In order to optimize the search for triples stored in the database, seven indexes are created automatically. %
Each index, stored on disk, is a data structure that contains all triples. %
Each combination of elements (represented by the order of the letters in an index name) indicates triples groups and order inside the files.
For example, the spogi index first classifies the subject (s), then predicate, object, graph and, finally, the id. %
The standard set of indexes is: spogi, posgi, ospgi, gspoi, gposi, gospi and i. %
From this initial configuration it is possible to create or remove indexes, for example, if there are no named graphs or this criterion is not used in queries, indexes starting with "g" will never be used and can be eliminated, optimizing the data load. %
Strings value like literal strings and IRIs are stored in a string table and the identifier, that references the respective string table entry, is stored in every triple index. %
Other values like integers, floats, booleans, dates, geospatial are encoded to efficient processing of range queries. %
The storage space needed per triple, when using the default indexes, is approximately 100 bytes. %

Some RDF Stores allow for attaching attributes to each triple as metadata augmentation in form of key-value pairs (keys refer to the attribute definitions) and, starting from version 6.1 (August, 2016), attributes, specified as key/value pairs in JSON format, can be associated with triples in the storage layer. %
It allows an arbitrary set of attributes to be defined per triple when the triple is created, that is, at the time of loading but cannot be changed afterwards. %
Triple attributes are not indexed and must be configured for the repository before inserting data. %
The potential applications suggested by its supplier for attributes include: (1) access control to restrict a user from accessing triples with certain attributes and (2) partitioning key to ensure that related triple are always stored together in the same shard. %
This attributes can also be used to: (1) add weights or costs to triples to be used in graph algorithms such as Dijkstra to finding the shortest paths between nodes in a graph, (2) automatic purging procedures can be implemented based on recorded time or expiration times of a triple and, (3) data quality dimensions such as accuracy, completeness, consistency, timeliness, validity, and uniqueness to represent veracity of Big Data Systems. %
We were able to test its RDF triple attributes implementation used to exemplify the transformation from LPG to RDF in \cite{besta2019demystifying}. %
Appendix \ref{sec:appendixA} provides examples of inserting and retrieving triples with attributes carried out in the tests.

Graph data stored in AllegroGraph is separated in repositories. %
Repositories are organized into catalogs. %
In its free version, it allows the storage of up to 5 million triples per repository (which can be used as a collection to group the correlated information in a single graph). %
There is always a root (/) catalog by default and a hidden system catalog, for internal use only. %
It is possible to create named catalogs <Catalog name>...</Catalog>) and dynamic catalogs (<DynamicCatalogs>...</DynamicCatalogs> ), the last one can be created and deleted through the HTTP interface while the server is running. %
Catalogs are locations on disk, specified in the configuration file, where its repositories are stored. %
Configuration parameters for named catalogs can also define some optional default settings for specific catalogs, like transaction log subdirectories, number and files size (TransactionLogDir, DesiredTlogFiles, TransactionLogSize), table compression rate (StringTableCompression), how often checkpoints will be performed (CheckpointInterval), and so on.  %
User permissions can be assigned to repository or catalog level. %

AllegroGraph main query language is SPARQL 1.1. %
Its SPARQL engine supports four query types for (sub)graph pattern matching: SELECT (project out specific variables and expressions, CONSTRUCT (construct RDF triples/graphs), ASK (ask whether or not there are any matches, result is either “true” or “false”.), and DESCRIBE (describe the resources matched by the given variables retrieving basic node/edge adjacency). %
Particularly, CONSTRUCT query form can be used for mapping subject, predicates and objects of stored triples to domain ontology's classes, relations and attributes. %
Although AllegroGraph do not provide features for view definition and maintenance, i.e., we cannot find any object that acts like a pre-established query command kept in the database dictionary\footnote{with AllegroGraph WebView the user can save a private copy of a query for later reuse}, for semantic graph databases, views could be useful for data accessing through SPARQL EndPoints to publish semantic data. %

In 2012 AllegroGraph was classified by \cite{renzo2012comparison} without support for Regular simple paths. %
AllegroGraph currently do supports basic graph path queries with fixed length paths (fixed number of nodes and edges) and regular simple paths (node and edge restrictions using regular expressions), using SPARQL Property Path Syntax. %
Appendix \ref{sec:appendixA} provides examples of patterning match and traversal queries carried out in the tests. %
It is important to mention that graph query capabilities are based on the query language not by the graph database itself. %
Considering \sparql{} language has no Data Definition clauses it is unknown why \cite {renzo2012comparison} indicated that AllegroGraph has Data Definition Language support. %
There is not a language pattern among graph databases but, considering the databases analyzed \cite{besta2019demystifying} we were able to see that all of the Triple Stores (10) support \sparql{}, as well as others databases categories such as OrientDB (Document), Oracle Spatial and Graph (RDBMS), OpenLink Virtuoso and Stardog (Data Hub). %

AllegroGraph interact with multiple programming languages and environments: Java, Python, Javascript, Lisp through API and other platforms using a RESTful HTTP protocol (using GET, PUT, POST). %
It supports reasoning through Prolog, a general purpose logic programming language, more suitable to write rules and express concepts that aren't "in triples". %
The same API can be used to connect to local, remote or federated triple stores. %
A federated triple store is an AllegroGraph feature to encapsulate multiple repositories into a single virtual store that can be manipulated as if it were one running parallel SPARQL query. %
Appendix \ref{sec:appendixA} provides examples of querying two repositories through federation carried out in the tests.

\begin{figure}
\centering
\includegraphics[width=0.8\textwidth]{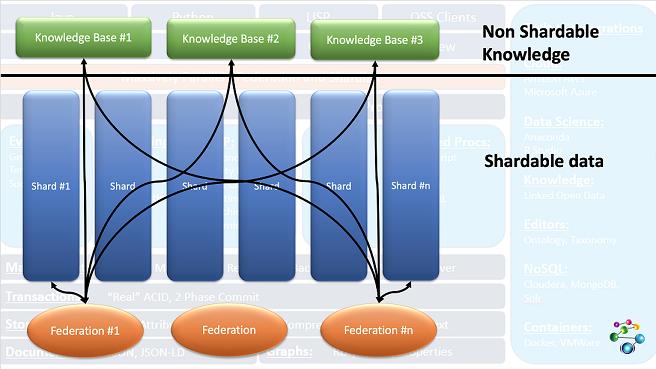}
\caption{AllegroGraph sharding}
\label{fig:fedshard}
\end{figure}

Parallel processing of queries in a single repository is not supported and, to provide horizontal scalability, AllegroGraph has sharding and replication features. %
Distributed Repositories is a collection of shards that are grouped together to store triples partitioned into shards across a cluster. %
The configuration of distributed repositories has a required partition key to be used when importing triples/quads. %
The key can be subject, predicate, object, or graph of a triple, or an attribute name. %
Queries can run in parallel but the data must be partitioned in such way  that each query runs independently on each shard, that is, without transferring data between them as seen in figure \ref{fig:fedshard}. %
\textit{Unshardable} data needed for all queries cannot be partitioned. %
Multi-master Replication mode can be used for standby, disaster recovery and load balancing. %
According to the documentation the replication solution is a real-time transactionally consistent data replication solution. %
Distributed systems has to sacrifice some availability in order to provide strong consistency, i. e., to guarantee data consistency across partitions and replicas. %
Based on this characteristics and CAP theorem definitions, we considered that AllegroGraph can be classified as CP (Consistency-Partition Tolerance) since it doesn't allow the configuration of consistency levels. %
It is different from \cite{persisting2017} where AllegroGraph was classified as AP (Availability-Partition Tolerance). %

\begin{figure}
\centering
\includegraphics[width=0.8\textwidth]{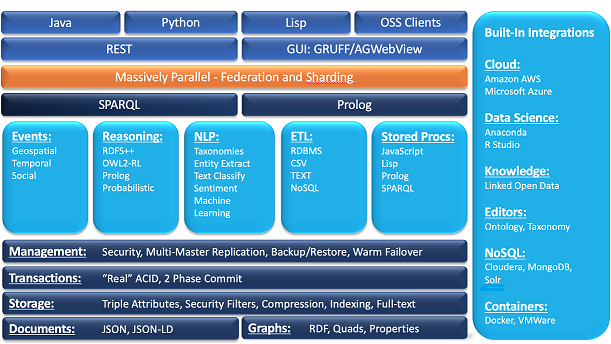}
\caption{AllegroGraph Components}
\label{fig:allegro}
\end{figure}

In addition to data storage, the product offers features of a database management system such as: (1) command line tool (agtool) for loading, exporting, querying, ..., (2) Web tool for administration ( AllegroGraph WebView) that allows the creation of users and roles, loading and exporting data, backup, query execution, index maintenance,…. (3) integration with third-party software such as Apache Solr, MongoDB, Anaconda, and ... (4) desktop tool (Gruff) to query, update, display graph stored in the repositories and also generate SPARQL and Prolog queries visually. %
AllegroGraph components can be seen in figure \ref{fig:allegro}. %

\section{Conclusion}
\label{sec:conclusions}

In this paper we gave an overview of NoSQL graph databases, focusing on the peculiarities that set them apart from other SQL and NoSQL databases. %
We also clarified how graph databases differ among themselves. %
Finally, we analyzed in-depth two different graph database management systems — AllegroGraph and Neo4j. %

\Myref{tab:allegro-neo4j-comparison} summarizes the characteristic features of the two systems, serving as a simple comparison between the two tools and as a reference for deciding on which system is better for a given use-case. %
Some of the characteristics listed in \myref{tab:allegro-neo4j-comparison} were not treated in the main text, either because of there is not much to be said about them than other than the information on the table (e.g., backups), or because of scope decisions (e.g., graph partitioning). %

The field of NoSQL (and in particular, graph databases) is fast-evolving, and has not stabilized just yet. %
We come to this conclusion by noticing that even recent papers like \cite{davoudian2018survey} have outdated information about particular systems, and many older papers discuss DBMS that have been deprecated or re-branded.\footnote{This has happened to all the systems discussed in \myref{sec:graphdbexamples}, for example.} %
There is still no clearly predominant query language for graph databases like SQL is relational databases, although the language being considered for standardization by ISO (see \myref{sec:graph-query}) might come to fulfill this role, if successful. %
There also no standard method or best practice for modeling data for graph DBMSs, notwithstanding \cite{akoka-four-vs-design}. %
To the best of our knowledge there are no studies verifying in practice the consequences of different models of the same data as graphs, nor the benefits and disadvantages of using schemas or not (in the systems that offer this possibility). %

There is also work to be done in bridging the academic and industrial worlds. %
Graph DBMSs (and this might extend to NoSQL and even all DBMSs) could be clear about the sort of consistency and isolation guarantees they offer, for example. %

As an example, \cite{berenson1995critique} notes that there are different interpretations to what the same isolation levels mean. %
Systems offering such guarantees should then clarify what definition they are using. %
But the fault is not only on the industrial actors: academia has long formalized isolation levels in a implementation-centric way, relying on behaviours that are invisible to their clients, and which and is difficult to reason about. %
\cite{crooks2017seeing} makes this critique and paves the way for client-centric definitions that are easier to observe, verify, and reason about. %

Another example where the gap between industry and academia could be breached is in graph partitioning. %
\cite[Appendix C]{davoudian2018survey} surveys graph partitioning algorithms and stores which implement them. %
There is a lot of work done in this area, but to the best of our knowledge none of the industrial-scale graph databases implement elaborate partitioning algorithms, and \cite[Appendix C]{davoudian2018survey} is of the same opinion. %

One area where there is also work to be done is in optimizing graph querying. %
Because graph pattern-matching is NP-complete \cite[§2.4]{davoudian2018survey}, optimal solutions to this problem are not practical, and thus heuristics and ad-hoc solutions abound. %
\cite[§5]{angles2017foundations} notes this, and encourages work to frame implementations together so that comparisons can be made. %
There might also be work on improving algorithms for special cases of graph pattern-matching (e.g., tree pattern-matching can be done in linear time \cite[§2.4]{davoudian2018survey}). %

\begin{table}[h]\footnotesize %
  \centering
  \begin{tabular}{||c||p{5cm}|p{5cm}||} %
\hline\hline
           \, &         \textbf{Allegro} & \textbf{Neo4j} \\\hline\hline
Data Model &          RDF & LPG\\\hline
Graph Storage       &Native & Native\\\hline
Data Representation & Triples, Quads & Edge/Adjacency lists\\\hline
Graph Processing    & Non-native, Index-based & Native\\\hline %

\textbf{Querying} & \multicolumn{2}{ c|| }{} \\\hline
Language & SPARQL & Cypher\\
QL Standardization  & W3C & openCypher\\
Query Pattern       & Pattern matching (full) Path/Navigational (some) & Pattern matching, Path/Navigational\\\hline %
Parallel Query   &   No (single repository)\newline{}Yes (federated repository) & Yes \\\hline %
\textbf{ACID} & \multicolumn{2}{ l|| }{} \\\hline
 Atomicity &       Yes (Rollback, Commit) & Yes\\
    Isolation       & snapshot isolation model without write concurrency guarantee & \textsc{read committed} (default), locks can be manually used for higher levels\\ %
    Durability &      Yes (Transaction Log) & Yes\\
    Consistency     & Strong Consistent without user-defined consistency rules & User-defined\\\hline
CAP &                CP (Consistent-Partition Tolerance) & AP (Causal consistency)\\\hline
Sharding &            Automatic (FedShard) & Manual\\\hline
Partition key &       subject, predicate, object, graph, triple attribute &  %
Manual\\\hline
Backup type         & Online full backups & Incremental online full backups\\\hline
Backup tools        & WebView and agtool backup & neo4j-admin\\\hline
Point-in-Time Recovery &  Yes (agtool recover) & No\\\hline
Replication model &   Multi-master & Multi-master\\\hline
Cloud-Hosted &        AWS or self-hosted & Neo4j Aura or self-hosted\\\hline %
Other indexes &       Full-text & Full-text\\\hline
Security Filters &    triple attribute & nodes, labels, properties, relationships, subgraphs\\\hline
User Permissions &   Catalog and Repository levels, Specific functions,     Roles & Role-based permissions, subgraph-access restrictions\\ %
\hline\hline
\end{tabular}

\caption{Overview of AllegroGraph and Neo4j}
\label{tab:allegro-neo4j-comparison}
\end{table}
\todo[disable]{na tabela, other indexes fica estranho sem mencionar os outros, não?} %


\bibliographystyle{abnt-puc}
\bibliography{refs}

\appendix{} %
\newpage{} %

\section{AllegroGraph tests in \sparql{}}
\label{sec:appendixA}

\begin{figure}
\centering
\includegraphics[width=1\textwidth]{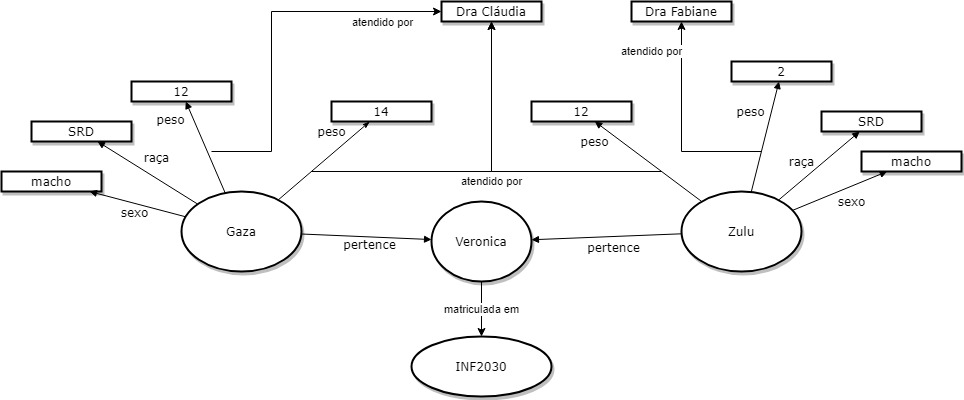}
\caption{Graph data example used in AllegroGraph tests}
\label{fig:graphexample}
\end{figure}

This appendix contains SPARQL commands used to manipulate (create and read) the graph data example of \myref{fig:graphexample} in AllegroGraph repositories.

\begin{lstlisting}
+++++++++++++++++++++++++++++++++++++++++++++++++++

# repository <inf2030-quad> used to store quads
# triples plus named graphs

INSERT DATA 
{
#triples
<Gaza>	<sex>	"macho" .
<Gaza>	<breed>	"SRD" . 
<Zulu>	<sex>	"macho" . 
<Zulu>	<breed>	"SRD" .

#quads, tagging data with its type
GRAPH <G3> {<Gaza>	<weight>	"14"^^xsd:decimal } . 
GRAPH <G3> {<Zulu>	<weight>	"12"^^xsd:decimal} .
GRAPH <G2> {<Zulu>	<weight>	"2"^^xsd:decimal} .
GRAPH <G1> {<Gaza>	<weight>	"10"^^xsd:decimal } . 

#named graphs and its attributes as triples
# tagging data with its type
<G1>	<served_by>	"Dra Claudia" .
<G1>	<served_in>	"2016-03-21"^^xsd:date .
<G2>	<served_by>	"Dra Fabiane" .
<G2>	<served_in>	"2017-03-08"^^xsd:date .
<G3>	<served_by>	"Dra Claudia" .
<G3>	<served_in>	"2019-06-10"^^xsd:date . 
}

# Retrieving triples of default graph and quads
# This PREFIX causes the default graph of the dataset to include
# only triples that are not in a named graph.
# Otherwise, the default graph will include every triple.
PREFIX franzOption_defaultDatasetBehavior: <franz:rdf>

# View quads - default graphs UNION named graphs
SELECT ?s ?p ?o ?g {
  # default graph
  { ?s ?p ?o . }
  UNION
  # named graphs
  { GRAPH ?g { ?s ?p ?o . }  }
}

# View unique named graphs
select distinct ?g { graph ?g { ?s ?p ?o } }

+++++++++++++++++++++++++++++++++++++++++++++++++++

# repository <inf2030-att> used to store triples with attributes

INSERT DATA 
{
<Gaza>	<sex>	"macho" .
<Gaza>	<breed>	"SRD" . 
<Zulu>	<sex>	"macho" . 
<Zulu>	<breed>	"SRD" .
}

# PREFIX is used to specify attributes with encoded JSON format 
# Without tagging data with its type

# Inserting triples with attributes 
# {"served_by": "Dra Claudia"}
# Attributes has to be configured in the repository before

PREFIX franzOption_defaultAttributes:
    <franz:%7B%22served_by%22%3A%20%22Dra%20Claudia%22%20%7D>
INSERT DATA { <Zulu>	<weight>	12  . }

# Inserting triples with attributes 
# {"served_by": "Dra Fabiane"}

PREFIX franzOption_defaultAttributes:
    <franz:%7B%22served_by%22%3A%20%22Dra%20Fabiane%22%20%7D>
INSERT DATA { <Zulu>	<weight>	2 .   }

# Inserting triples with attributes 
# {"served_by": "Dra Claudia"}

PREFIX franzOption_defaultAttributes: 
    <franz:%7B%22served_by%22%3A%20%22Dra%20Claudia%22%20%7D>
INSERT DATA { <Gaza>	<weight>	14 . <Gaza>	<weight>	10 . }

# Retrieving triples and its attributes 

select distinct ?s ?p ?o ?key ?value 
{ (?key ?value) 
  <http://franz.com/ns/allegrograph/6.2.0/attributesNameValue> 
  (?s ?p ?o) }

+++++++++++++++++++++++++++++++++++++++++++++++++++

# Federation with two local repositories 
# <inf2030-att>+<inf2030-quad>
# Retrieving unique triples - DISTINCT 
# Ordered by predicate - ORDER BY

SELECT DISTINCT ?s ?p ?o { ?s ?p ?o . } ORDER BY DESC(?p)

+++++++++++++++++++++++++++++++++++++++++++++++++++

# (Sub)Graph Pattern Matching

# Filtering triples with predicate equal <sex>

SELECT ?s ?o  {?s <sex> ?o}

# Changing local predicate by an ontology property

PREFIX sch: <http://schema.org#>
CONSTRUCT {?s sch:gender ?o.} where {?s <sex> ?o}
CONSTRUCT {?s sch:weight ?o.} where {?s <weight> ?o}

# Checking if there is any triple with predicate equal <idade>

ASK {?s <idade> ?o}

# Getting all neighbours of a node
# Subject with predicate equal <breed>

DESCRIBE ?s WHERE {?s <breed> ?o}

+++++++++++++++++++++++++++++++++++++++++++++++++++

# Graph Path/Traversal Queries

INSERT DATA 
{
<Gaza>	<belongs_to>	<Veronica> .
<Zulu>	<belongs_to>	<Veronica> .
<Veronica>	<enrolled_in>	<INF2030> .
}

# Regular simple paths

ASK { <Gaza> (<belongs_to>* / ^<belongs_to>*) <Zulu> }

# Fixed length paths

ASK { <Gaza> (<belongs_to> / <enrolled_in>) <INF2030> }

\end{lstlisting}

\end{document}